\documentclass[aps,preprint,english,superscriptaddress,citeautoscript,showkeys,preprintnumbers,amsmath,amssymb,floatfix,footinbib,prb]{revtex4-2}
\usepackage{hyperref}
\hypersetup{colorlinks=true,linkcolor=red,citecolor=blue}
\usepackage{dcolumn}
\usepackage{float}
\usepackage{soul}
\usepackage{graphicx,bm}
\usepackage{amssymb,amsmath}
\usepackage{txfonts}

\DeclareUnicodeCharacter{FB01}{fi}

\begin{document}

\title{Anisotropic superconductivity and unusually robust electronic critical field in single crystal La$_{7}$Ir$_{3}$}

\author{D. A. Mayoh}
\email[]{d.mayoh.1@warwick.ac.uk}
\affiliation{Physics Department, University of Warwick, Coventry, CV4 7AL, United Kingdom}

\author{S. J. R. Holt}
\affiliation{Physics Department, University of Warwick, Coventry, CV4 7AL, United Kingdom}

\author{T. Takabatake}
\affiliation{Graduate School of Advanced Science and Engineering, Hiroshima University, Higashi-Hiroshima 739-8530, Japan}

\author{G. Balakrishnan}
\affiliation{Physics Department, University of Warwick, Coventry, CV4 7AL, United Kingdom}

\author{M. R. Lees}
\email[]{m.r.lees@warwick.ac.uk}
\affiliation{Physics Department, University of Warwick, Coventry, CV4 7AL, United Kingdom}

\begin{abstract}

Polycrystalline La$_{7}$Ir$_{3}$ is reported to show superconductivity breaking time-reversal symmetry while also having an isotropic $s$-wave gap. Single crystals of this noncentrosymmetric superconductor are highly desirable to understand the nature of the electron pairing mechanism in this system. Here we report the growth of high-quality single crystals of La$_{7}$Ir$_{3}$ by the Czochralski method. The structural and superconducting properties of these large crystals have been investigated using x-rays, magnetization, resistivity and heat capacity measurements. We observe a clear anisotropy in the lower and upper critical fields for magnetic fields applied parallel and perpendicular to the hexagonal $c$ axis. We also report the presence of a robust electronic critical field, that diverges from the upper critical field derived from heat capacity, which is the hallmark of surface superconductivity.

\end{abstract}

\maketitle

\section{Introduction}

Noncentrosymmetric (NCS) superconductors have attracted a great deal of experimental and theoretical interest in recent years due to the range of exotic superconducting properties they exhibit~\cite{Smidman17, Bauer12}. The lack of inversion symmetry in these materials means that parity is no longer a good quantum number and antisymmetric spin-orbit coupling can lead to an admixture of singlet and triplet Cooper pairs. While it is usual for one channel to dominate, such an admixture of singlet and triplet pairs may result in novel superconducting phenomena, including unusual superconducting gap structures~\cite{Nishiyama2007,Yuan2006,Mayoh18}, unconventional upper-critical field behavior~\cite{Mayoh19}, and time-reversal symmetry breaking (TRSB) ~\cite{LaNiC2,LaNiGa2,Re6Zr, Re6Hfc, Shang18, Re6Ti, Shang18a, La7Ir3,La7Rh3,Mayoh21,La7Ni3TRSB}. 

Time-reversal symmetry breaking in NCS superconductors is a rare, poorly understood phenomena, and to date has been reported in relatively few classes of NCS superconductors. For example, TRSB has been observed in LaNiC$_{2}$ (and its centrosymmetric (CS) analogue LaNiGa$_{2}$)~\cite{LaNiC2,LaNiGa2}, some rhenium based materials with an $\alpha$-Mn structure~\cite{Re6Zr, Re6Hfc, Shang18, Re6Ti, Shang18a}, tetragonal $\alpha$-V$_3$S-type Zr$_3$Ir~\cite{ShangZr3Ir}, and the hexagonal La$_{7}X_{3}$ compounds where $X$ is Ir, Rh, Pd or Ni~\cite{La7Ir3,La7Rh3,Mayoh21,La7Ni3TRSB}. Further characterization has indicated largely conventional, nodeless superconducting properties for all these groups of materials~\cite{Weng2016,Re6Zr,Khan16,Mayoh17,La7Ir3,Li17,ShangZr3Ir,SajileshZr3Ir}. These results are puzzling. In LaNiC$_{2}$ and LaNiGa$_{2}$, a nodeless two-gap superconducting state has been suggested to account for the contradictory behavior~\cite{Weng2016}, while the signature of TRSB has been reported in pure rhenium, suggesting that factors beyond the NCS structure may need to be considered to explain the superconducting properties of the Re based materials~\cite{Shang18a}. For Zr$_3$Ir there are other reports that TRS is preserved~\cite{SajileshZr3Ir}. Much uncertainty also remains over the origin of TRSB in La$_{7}$Ir$_{3}$ and related compounds. Polycrystalline averaging makes it experimentally difficult to distinguish subtle changes in the superconducting properties that should be the hallmark of singlet-triplet mixing and to date there have been no studies on single crystals of La$_{7}$Ir$_{3}$. Single crystals of these compounds are essential in order to carry out detailed investigations of any anisotropy in the superconducting properties, in order to reveal the true nature of the superconducting state of La$_{7}$Ir$_{3}$ and the related La$_{7}X_{3}$ compounds. Furthermore, the high degree of tunability of the spin orbit coupling of the La$_{7}X_{3}$ family makes them excellent candidates for studying the effects of the strength of the spin-orbit coupling on the observed unconventional superconducting properties.

Here we report the first successful growth of large single crystals of the NCS superconductor La$_{7}$Ir$_{3}$ by the Czochralski method. The crystals obtained are large enough to perform both laboratory based physical property measurements, as well as neutron scattering and muon spectroscopy experiments. We demonstrate the high-quality of the crystals obtained using x-ray diffraction techniques and compositional analysis. We also report on the magnetic, transport, and thermodynamic properties of La$_{7}$Ir$_{3}$ and discuss the anisotropy of the superconducting properties of this important superconducting material.

\section{Single Crystal Growth and Structural Characterization}

The binary phase diagram of La-Ir indicates that La$_{7}$Ir$_{3}$ forms congruently at $1140~^{\circ}$C~\cite{Massalski86} which makes it an excellent candidate for single crystal growth by the Czochralski method.

\subsection{Synthesis of polycrystalline La$_{7}$Ir$_{3}$ precursor}

A polycrystalline button of La$_{7}$Ir$_{3}$ was made using stoichiometric amounts of elemental La (Ames Laboratory, 99.9\%) and Ir (Goodfellow Metals, 99.9\%). These metals were melted together in a Hukin-type cold crucible using radio frequency (RF) induction heating under a high purity argon atmosphere. The sample button was flipped and remelted to improve the homogeneity of the sample. The observed weight loss during the melting was negligible. 

\subsection{Single crystal growth}
Single crystals of La$_{7}$Ir$_{3}$ were grown using a modified Crystalox system with a pulling mechanism. The sample chamber was flushed twice with high purity argon and evacuated to a vacuum of $1.4 \times 10^{-6}$~mbar. The La$_{7}$Ir$_{3}$ sample was melted in a tungsten crucible and a single crystal was grown by the Czochralski method using a tungsten seed rod at an initial pulling rate 6~mm/hr in an argon atmosphere. The tungsten rod was rotated at 7~rpm.
 
\begin{figure}[t]
\centering
\includegraphics[width=0.9\columnwidth]{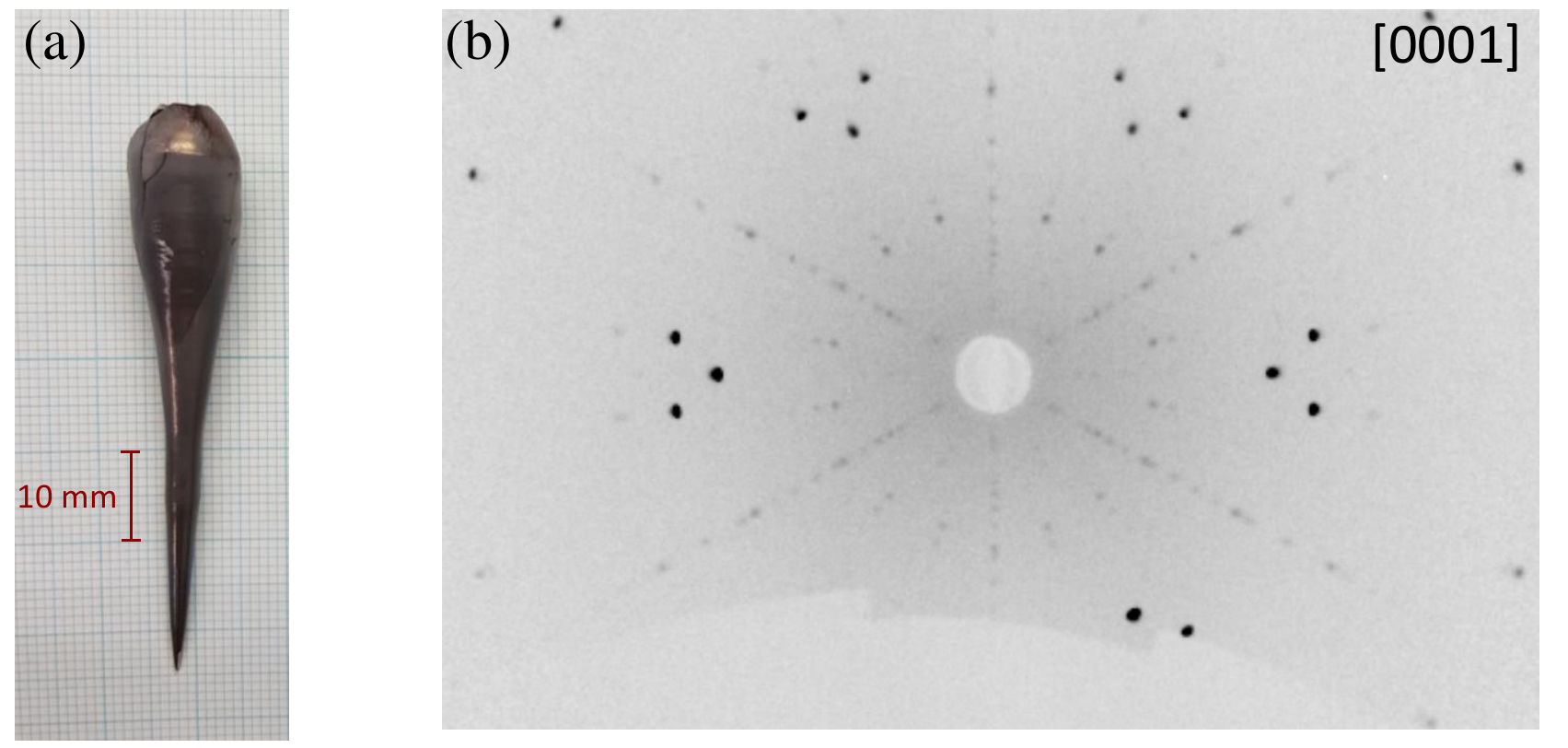}
\caption{(Color online) (a) As grown crystal boule of La$_{7}$Ir$_{3}$ obtained by the Czochralski process. (b) X-ray Laue diffraction pattern of an isolated section of the La$_{7}$Ir$_{3}$ boule orientated along the $[0001]$ direction and used for physical property measurements.}
\label{FIG: La7Ir3 crystal}
\end{figure}

A crystal of La$_{7}$Ir$_{3}$ approximately 60~mm in length with a maximum diameter of around 10~mm was successfully obtained as shown in Fig~\ref{FIG: La7Ir3 crystal}(a). The crystal boule was not annealed after the growth. The boule was seen to tarnish and discolor quickly indicating that the sample is air sensitive. 
 
\subsection{X-ray Laue}
A Photonic-Science Laue camera system was used to record back-scattered x-ray Laue diffraction patterns of the sample. A triple axis goniometer allowed for the single crystals to be orientated along high symmetry directions.
 
The entire length of the as-grown boule of La$_{7}$Ir$_{3}$ was examined to confirm its quality. The sharp spots obtained in all the patterns confirmed the sample has a high degree of crystallinity. Small sections of crystal were cut with faces perpendicular to either the $c$ axis ($[0001]$ direction) or the $ab$ plane ($[2\bar{1}\bar{1}0]$ direction) for further study. The Laue diffraction pattern of an isolated section of the La$_{7}$Ir$_{3}$ boule, orientated along the $[0001]$ direction and showing the expected six-fold symmetry, is shown in Fig.~\ref{FIG: La7Ir3 crystal}(b). 
 
\subsection{Powder x-ray diffraction}

\begin{figure}[t]
\centering
\includegraphics[width=0.7\columnwidth]{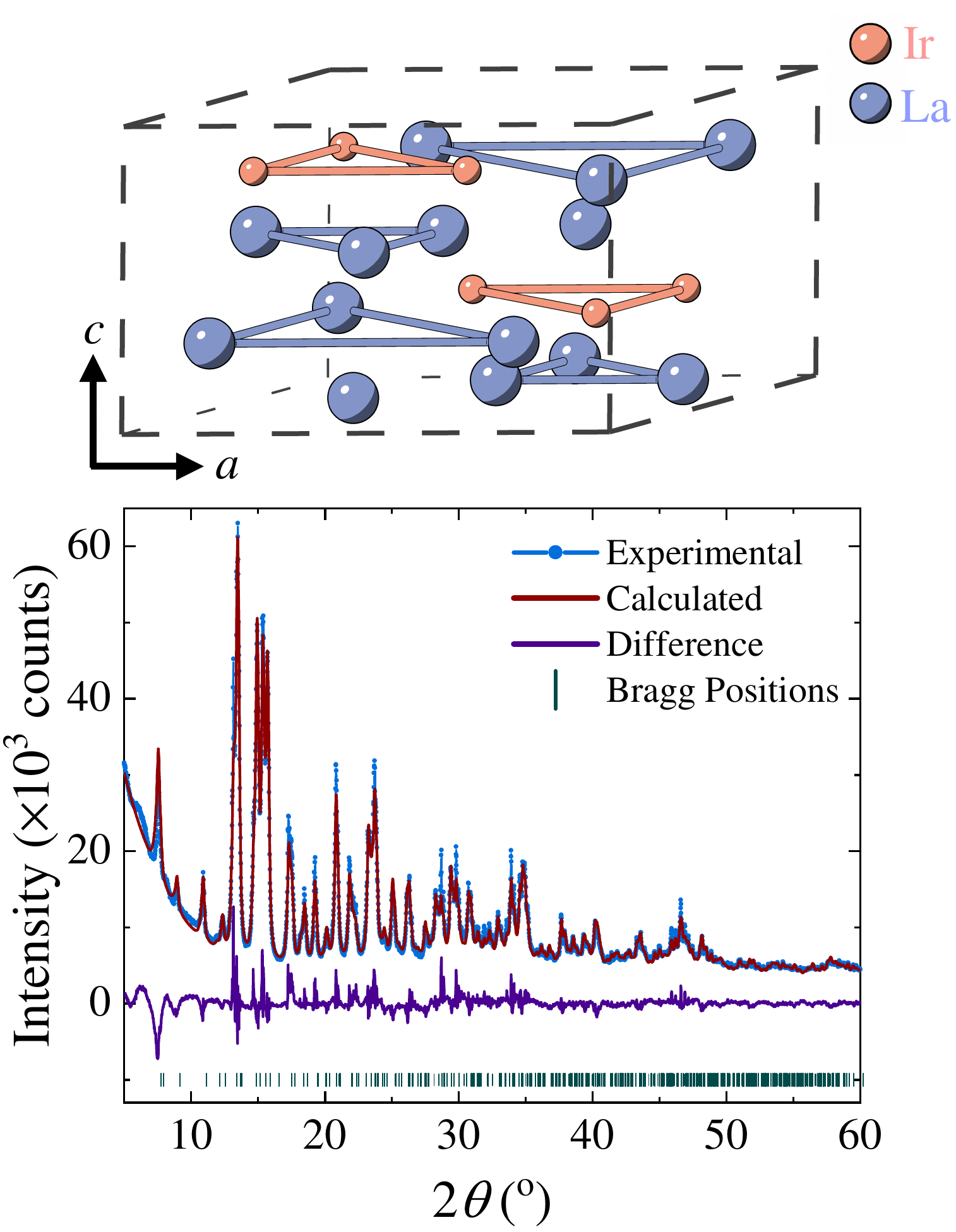}
\caption{(Color online) Powder x-ray (Mo K$\alpha$) diffraction pattern collected at 300~K on a ground piece of the La$_{7}$Ir$_{3}$ boule. The blue points show the x-ray pattern, the red line shows the Rietveld refinement to the diffraction pattern (space group $P 6_{3} mc$ - No. 186), and the purple line shows the difference between the fit and the data. The expected Bragg peaks are indicated by the green dashes. The unit cell of La$_{7}$Ir$_{3}$ is shown where the blue and red spheres represent La and Ir atoms, respectively.}
\label{FIG: La7Ir3 pxrd}
\end{figure}

A small section of the as-grown boule was ground and examined via powder x-ray diffraction using a Panalytical Empyrean diffractometer equipped with a Mo target and an Oxford Cryosystems PheniX sample chamber to reduce the oxidation of the La$_{7}$Ir$_{3}$ powder. The results of the measurements can be seen in Fig.~\ref{FIG: La7Ir3 pxrd}. La$_{7}$Ir$_{3}$ crystallizes in the NCS hexagonal Th$_{7}$Fe$_{3}$-type, (space group $P6_{3}mc$ - No. 186) crystal structure. The refinement of the diffraction pattern by Rietveld analysis using the TOPAS-academic software package~\cite{Coelho:jo5037} yielded the correct $P 6_{3} mc$ hexagonal structure, with lattice parameters $a = 10.2382(5)$~\AA\ and  $c = 6.4778(5)$~\AA\, in strong agreement with previous reports~\cite{Li17,Geballe65}. No impurities can be seen to within the resolution limit of the instrument. The crystallographic parameters obtained from the refinement are given in Table.~\ref{Tab:La7Ir3 pxrd}.

\subsection{Energy-dispersive x-ray spectroscopy}
A ZEISS GeminiSEM 500 was used to perform energy-dispersive x-ray spectroscopy (EDX) measurements on an isolated piece of the La$_{7}$Ir$_{3}$ boule, to check the stoichiometry and search for any elemental variations across the sample. Measurements were made averaging across the bulk of the sample as well as at six selected sites on the crystal surface. The average stoichiometry was measured to be 70.5(2)\% La and 29.5(2)\% Ir. Measurements at the individual sites showed no significant local variations in stoichiometry across the specimen studied. The largest variation was a 0.7\% increase in La observed at one site giving a stoichiometry 71.2\% La and 28.8\% Ir. A typical EDX spectrum and a scanning electron microscope image of the sample surface are provided in the Supplemental Material~\cite{SMnote}.
 
%%%%%%%%%%%%%%%%%%%%%%%%%%%%%%%%%%%%%%%%%%

\section{Superconducting Magnetic and Transport Properties}

Single crystals of La$_{7}$Ir$_{3}$ provide an excellent opportunity to look for any anisotropic behavior that may explain the unusual superconducting properties previously reported in polycrystalline samples~\cite{La7Ir3,Li17}. In the following section the magnetic, electrical, and thermal properties of the high-quality single crystals of La$_{7}$Ir$_{3}$ were investigated in both the normal and superconducting state. 

\subsection{Magnetization and lower critical field}

\begin{figure*}[t]
\centering
\includegraphics[width=\columnwidth]{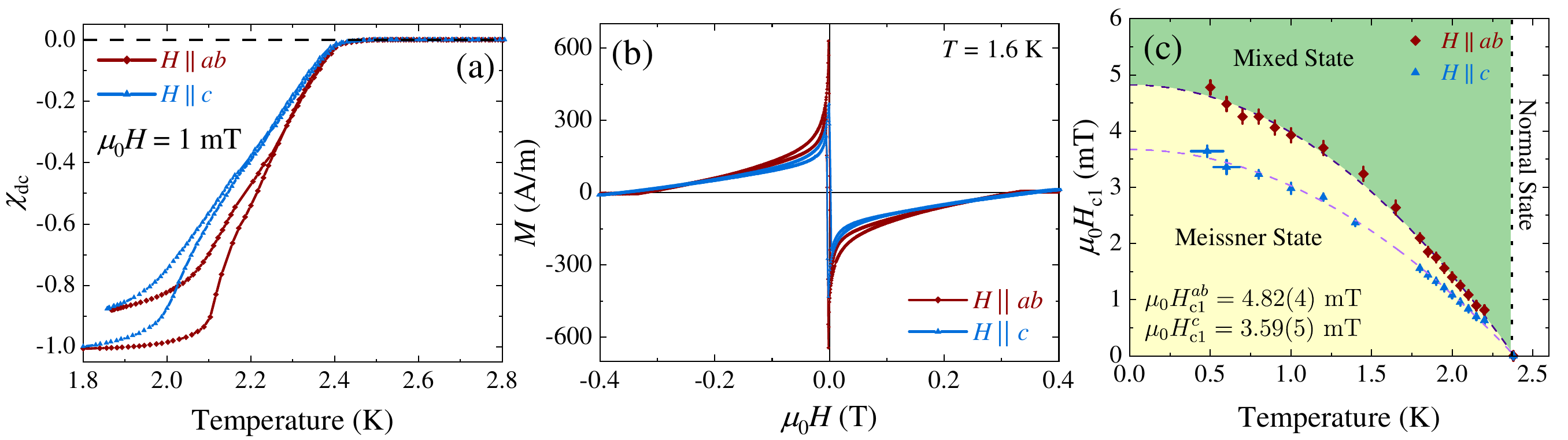}
\caption{(Color online) (a) Temperature dependence of the magnetic susceptibility for La$_{7}$Ir$_{3}$ with an applied magnetic field of 1~mT along the $c$ axis ($[0001]$ direction) (blue) or in the $ab$ plane ($[2\bar{1}\bar{1}0]$ direction) (red). The samples were measured in the ZFCW and FCC modes. (b) Field dependence of the magnetization for La$_{7}$Ir$_{3}$ at 1.6~K along the $c$~axis (blue) and in the $ab$~plane (red) (c) Lower critical field as a function of temperature for the $H \parallel c$ axis (blue) and $H \parallel ab$ plane (red) crystal directions. The dashed lines show fits to the data using the Ginzburg-Landau expression $H_{\mathrm{c1}}\left(T\right) = H_{\mathrm{c1}}\left(0\right)\left[1 - \left(T/T_{\mathrm{c}}\right)^{2}\right]$.}
\label{FIG: La7Ir3 Magnetisation}
\end{figure*}

Samples cut from the as-grown boule along the desired crystallographic orientations were investigated using magnetization. A Quantum Design (QD) Magnetic Property Measurement System (MPMS) equipped with an i-Quantum He-3 insert allowed dc magnetization measurements to be performed at temperatures between 0.5 to 300~K in applied fields up to 5~T. The temperature dependence of the dc magnetic susceptibility in a field  of 1~mT applied along the $c$ axis and in the $ab$ plane are shown in Fig.~\ref{FIG: La7Ir3 Magnetisation}(a). Additional measurements made in higher applied fields are presented in the Supplemental Material~\cite{SMnote}. The samples were measured in zero-field-cooled-warming (ZFCW) and field-cooled-cooling (FCC) mode between 1.8 and 2.8~K. The magnetic susceptibility data were corrected for the demagnetization factors determined from the sample shapes~\cite{Aharoni1998}. The onset of superconductivity is observed at $T_{\mathrm{c}}^{\mathrm{onset}} = 2.41(5)$~K, which is higher than the $T_{\mathrm{c}} = 2.25(5)$~K reported in polycrystalline La$_{7}$Ir$_{3}$~\cite{La7Ir3,Li17}. The increase in $T_{\mathrm{c}}$ may be attributed to the lack of strain and disorder in the single crystal. Immediately below $T_{\mathrm{c}}$, the ZFCW and FCC curves overlap, as the sample is in the mixed state with weak pinning. Below $H_{\mathrm{c1}}\left(T\right)$ the signal changes more rapidly. A full Meissner fraction of $\chi_{\mathrm{dc}}^\mathrm{ZFCW} = -1$ is observed for fields applied along both crystallographic directions indicating bulk superconductivity in these single crystals of La$_{7}$Ir$_{3}$. When the sample is cooled in field, a significant fraction of the field is re-excluded with $\chi_{\mathrm{dc}}^\mathrm{FCC} = -0.9$ in both directions. This emphasizes the defect-free nature of the crystals. At higher magnetic fields, the $M(H)$ loops collected in an Oxford Instruments vibrating sample magnetometer are almost reversible up to $H_{\mathrm{c2}}\left(T\right)$ as shown in Fig.~\ref{FIG: La7Ir3 Magnetisation}(b). This is further evidence of the defect-free nature of the crystals with weak pinning.  Measurements of the ac susceptibility versus temperature of La$_7$Ir$_3$ in different applied fields~\cite{SMnote} are consistent with these observations. 

%%%% lower critical field

The lower critical fields, $H_{\mathrm{c1}}\left(T\right)$, for fields along the $c$ axis and in the $ab$ plane were determined from the magnetization versus applied field curves collected at fixed temperature. $H_{\mathrm{c1}}$ is taken to be the field  at which the magnetization versus field curves first deviate from linearity~\cite{Umezawa88}. Figure~\ref{FIG: La7Ir3 Magnetisation}(c) shows $H_{\mathrm{c1}}\left(T\right)$ for both directions. The data have been fit using the Ginzburg-Landau (GL) expression, $H_{\mathrm{c1}}\left(T\right) = H_{\mathrm{c1}}\left(0\right)\left[1 - \left(T/T_{\mathrm{c}}\right)^{2}\right]$ giving $\mu_{0}H_{\mathrm{c1}}^{ab}(0) = 4.82(4)$~mT in the $ab$ plane and $\mu_{0}H_{\mathrm{c1}}^c(0) = 3.59(5)$~mT along $c$. This anisotropy of the lower critical field values may indicate the presence of a subtle anisotropy in the superfluid density of La$_{7}$Ir$_{3}$. However, further studies to probe the anisotropy of the penetration depth using methods such as muon spectroscopy or tunnel-diode oscillator measurements are necessary to confirm this.

\subsection{Electrical resistivity}

\begin{figure}[t]
\centering
\includegraphics[width=0.6\columnwidth]{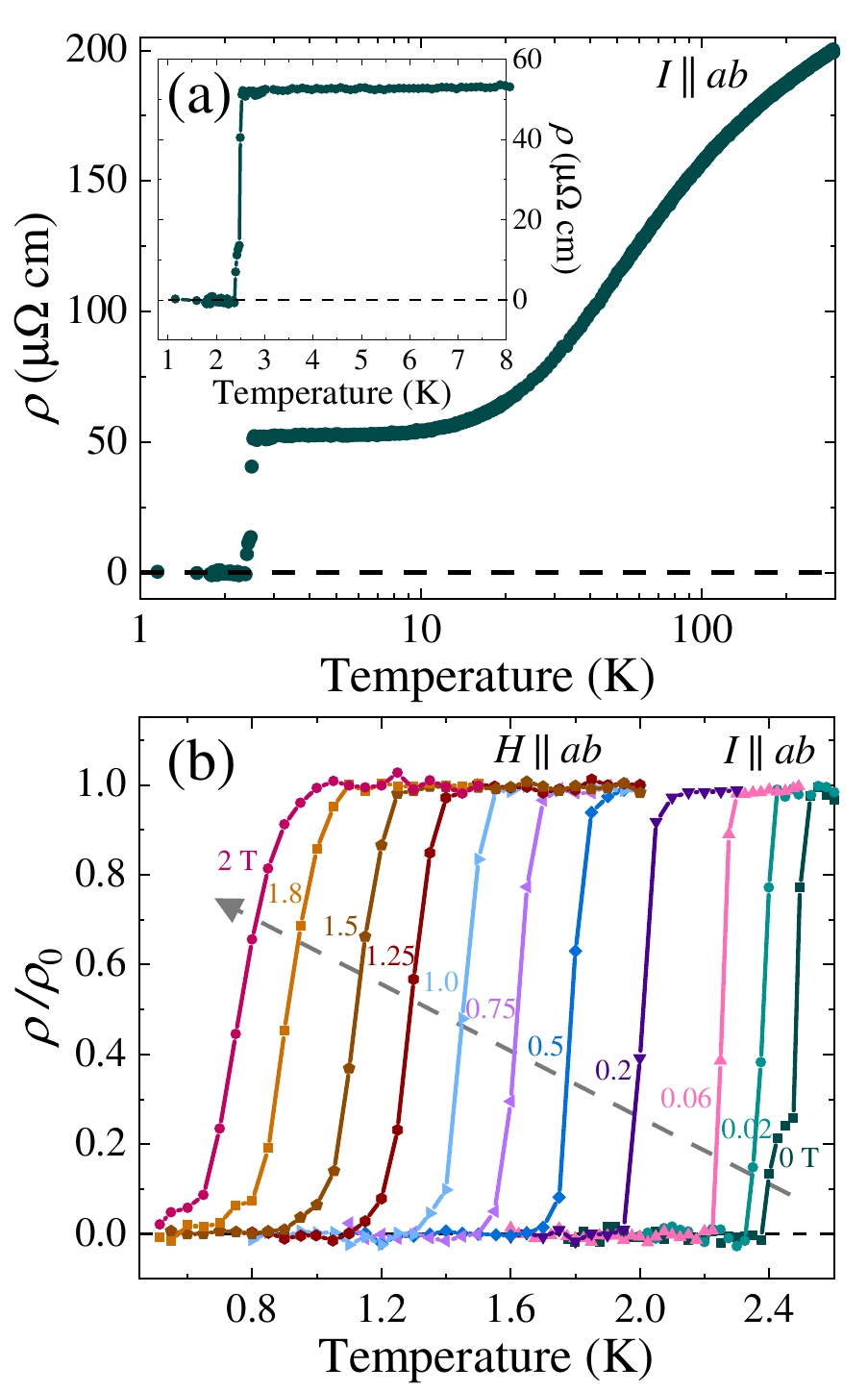}
\caption{(Color online) (a) Temperature dependence of the superconducting and normal-state resistivity for La$_{7}$Ir$_{3}$ between 1.8 and 300~K with the measuring current applied along the $[\bar{1}2\bar{1}0]$ crystal direction. La$_{7}$Ir$_{3}$ has a superconducting transition at $T_{\mathrm{c}} = 2.38(5)$~K. The inset shows the temperature dependence of the electrical resistivity between 1 and 8~K. (b) Electrical resistivity versus temperature for selected applied fields between 0 and 2~T. The field was applied along the $[2\bar{1}\bar{1}0]$ direction and the current passed along the $[\bar{1}2\bar{1}0]$ direction, within the $ab$~plane. The superconducting transition is seen to be suppressed in temperature and broadened for increasing fields.}
\label{FIG: La7Ir3 Resistivity}
\end{figure}

Electrical resistivity measurements were carried out on a bar shaped sample of La$_{7}$Ir$_{3}$ to ensure a uniform electropotential across the sample. Measurements were performed using a Quantum Design Physical Property Measurement System (PPMS) with a He-3 insert allowing measurements to be performed between 0.5 and 300~K in fields up to 2~T. The temperature dependence of the electrical resistivity from 1 to 300~K with the current applied along the $[\bar{1}2\bar{1}0]$ crystallographic direction can be seen in Fig.~\ref{FIG: La7Ir3 Resistivity}(a). The normal-state resistivity data can be a strong indicator of the quality of the crystal. The crystals measured have a residual resistivity $\rho_{0} = 52.5(5)~\mathrm{\mu\Omega}$~cm and a residual resistivity ratio (RRR) of 3.8, which is slightly higher than the RRR of 3.4 previously reported in polycrystalline samples. Using a simple free electron model, the mean free path is estimated from the residual resistivity to be 16(1)~nm, which is similar to the coherence length (see Sec.~\ref{UpperHc2}), placing La$_{7}$Ir$_{3}$ in the dirty limit. The high crystallinity and weak pinning suggest the residual resistivity is not entirely due to impurity or defect scattering.

A superconducting transition is observed at $T_{\mathrm{c}} = 2.47(5)$~K, as shown in Fig.~\ref{FIG: La7Ir3 Resistivity}(a), indicating the onset of superconductivity and above this temperature no further transitions are seen.  The temperature dependence of the electrical resistivity in several fields applied in the $ab$~plane is shown in Fig.~\ref{FIG: La7Ir3 Resistivity}(b). Resistivity versus temperature curves for fields applied along the $c$ axis are given in the Supplemental Material~\cite{SMnote}. In 20~mT, a single sharp superconducting transition ($\Delta T = 0.07$~K) is observed. However, in fields below 20~mT a shoulder in the transition is observed. No shoulder is observed at the superconducting transition in the magnetization or heat capacity data. Figure~\ref{FIG: La7Ir3 Resistivity}(b) shows how the superconducting transition for La$_{7}$Ir$_{3}$ is suppressed and broadened for applied fields from 0.2 to 2~T when the field is applied in the $ab$~plane. At 1.7~T, $T_{\mathrm{c}} = 0.98(5)$~K with $\Delta T = 0.5$~K. The rounding of the transition below $T_{\mathrm{c}}^{\mathrm{onset}}$ may be indicative of surface superconductivity (see Sections~\ref{HeatC} and~\ref{UpperHc2} below) as a superconducting path is established more easily, with bulk superconductivity only achieved at lower temperatures~\cite{Zeinali16}. The broadening of the transition closer to zero resistivity may be the result of flux motion, in these single crystals with relatively weak pinning. 

\subsection{Heat capacity}
\label{HeatC}
\begin{figure*}[t]
\centering
\includegraphics[width=\columnwidth]{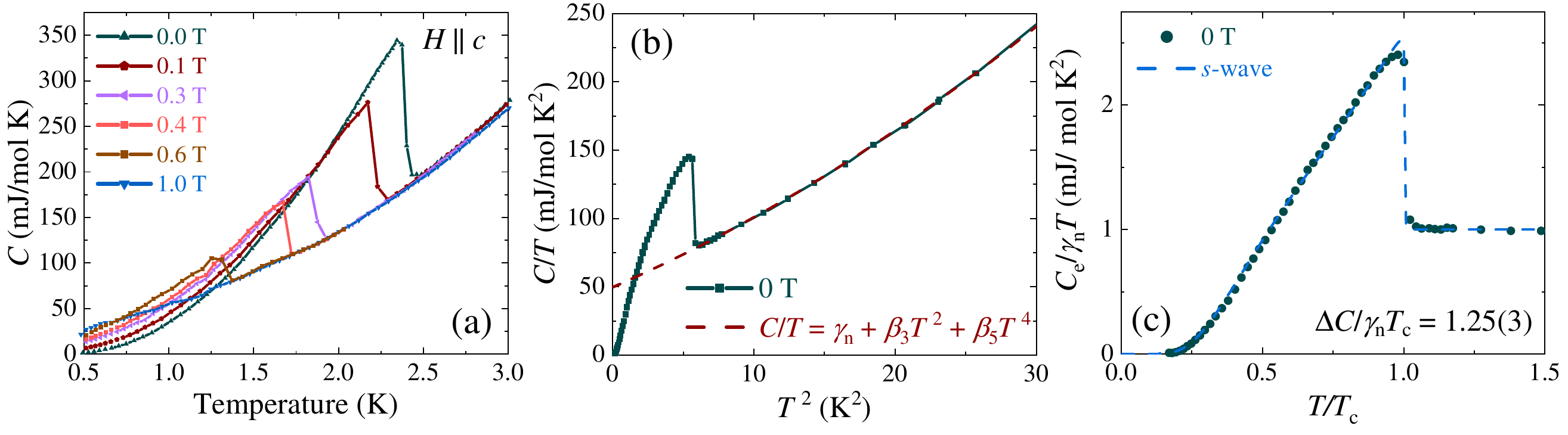}
\caption{(Color online) (a) Temperature dependence of the heat capacity for La$_{7}$Ir$_{3}$ in several fields up to 1 T applied along the $c$ axis. (b) $C/T$ versus $T^{2}$ for La$_{7}$Ir$_{3}$ below 5 K in zero-field. The dash line shows the fit to data using $C/T = \gamma_{\mathrm{n}} + \beta_{3}T^{2} + \beta_{5}T^{4}$ and its extrapolation to $T=0$~K. (c) Temperature dependence of the normalized heat capacity $C_{e}/\gamma_{\mathrm{n}}T$. The dashed line indicates the fit to the data using an $s$-wave superconducting gap model described in the main text.}
\label{FIG: La7Ir3 HC}
\end{figure*}

In order to explore the nature of the superconducting pairing mechanism in La$_{7}$Ir$_{3}$, measurements of the heat capacity were carried out. Measurements were made in a QD-PPMS using a two-tau relaxation method. The temperature dependence of the heat capacity in magnetic fields up to 1~T applied along the $c$~axis are shown in Fig.~\ref{FIG: La7Ir3 HC}(a).  Similar measurements, (not shown), were made with the magnetic field applied in the $ab$ plane. In zero-field, a single sharp, $(\Delta T = 0.06~\mathrm{K})$, superconducting transition can be observed at $T_{\mathrm{c}} =2.37(5)$~K. The superconducting transition is suppressed with increasing field, however, the transition remains sharp indicating the high quality and homogeneity of our crystals.

The normal-state contribution to the specific heat of La$_{7}$Ir$_{3}$ can be obtained by fitting the data above the superconducting transition to $C/T = \gamma_{\mathrm{n}} + \beta_{3}T^{2} + \beta_{5}T^{4}$ as shown in Fig.~\ref{FIG: La7Ir3 HC}(b). Here $\gamma_{\mathrm{n}}$ is the Sommerfeld coefficient, $\beta_{3}$ is the Debye model lattice contribution and $\beta_{5}$ accounts for higher-order lattice contributions. The fit to the data gives $\gamma_{\mathrm{n}} = 49.6(5)$~mJ/mol K$^2$ and $\beta_{3} = 4.50(6)$~mJ/mol K$^4$ giving a Debye temperature $\theta_{\mathrm{D}}=\left(163 \pm 1 \right)$~K. These values are consistent with those reported for polycrystalline samples~\cite{Li17}. It has been mentioned that the value of $\gamma_{\mathrm{n}}$ suggests an enhanced density of states. Note, however, that $\gamma_{\mathrm{n}}/\mathrm{La} = 7.09(7)$~mJ/mol K$^2$ which is lower that the $\gamma_{\mathrm{n}} = 11.3$~mJ/mol K$^2$ of pure La~\cite{HeinigerLa1973}.

The temperature dependence of the zero-field electronic heat capacity $C_{\mathrm{e}}\left(T\right)$ can provide information on the structure of the superconducting gap and the nature of the pairing mechanism. The data are well fit using the Bardeen-Cooper-Schrieffer (BCS) model of specific heat with a single, isotropic $s$-wave gap. The specific heat in the superconducting state can be related to the entropy, $S_{\mathrm{e}}$, by $\dfrac{C_{\mathrm{e}}}{\gamma_{\mathrm{n}}T}=T\dfrac{d(S_{\mathrm{e}}/\gamma_{\mathrm{n}} T_{\mathrm{c}})}{dT}.$ $S_{\mathrm{e}}$, was calculated using $\dfrac{S_{\mathrm{e}}}{\gamma_{\mathrm{n}}T_{\mathrm{c}}}=-\dfrac{6}{\pi^2}\dfrac{\Delta\left(0\right)}{k_{\mathrm{B}}T_{\mathrm{c}}}\int_{0}^{\infty}\left[f\mathrm{ln}f+\left(1-f\right)\mathrm{ln}\left(1-f\right)\right]d\epsilon,$ where $\Delta(0)$ is the superconducting gap at zero temperature, $f\left(E\right)$ is the Fermi-Dirac function given by $f\left(E\right) = \left[1+\exp\left(E/k_{\mathrm{B}}T\right)\right]^{-1}$, $k_{\mathrm{B}}$ is the Boltzmann constant, and ${E=\sqrt{\epsilon^2+\left[\Delta\left(T\right)\right]^{2}}}$. Here $\epsilon$ is the energy of the normal state electrons and $\Delta\left(T\right)$ is the temperature dependence of the superconducting gap which is approximated using $\Delta\left(T\right) = \Delta(0)\tanh\lbrace 1.82\left[1.018\left(T_{\mathrm{c}}/T-1\right)\right]^{0.51}\rbrace$~\cite{Carrington}. The resulting fit to the data is shown in Fig.~\ref{FIG: La7Ir3 HC}(c) with $\Delta(0)/k_{\mathrm{B}}T_{\mathrm{c}} = 1.80(4)$ which is larger than the BCS weak coupling value (cf. 1.74) indicating a small increase in the electron-phonon coupling strength. 

The thermodynamic critical field, $H_{\mathrm{c}}$, can be determined from the zero-field heat capacity by calculating the difference between the free energy per unit volume of the normal and superconducting states, $\Delta F = \int_{T_{\mathrm{c}}}^{T} \int_{T_{\mathrm{c}}}^{T'} (C_{\mathrm{sc}} - C_{\mathrm{n}})/T'' \,dT'' dT' = H_{c}^{\mathrm{HC}}(T)^{2}/8 \pi$ giving $\mu_0H_{\mathrm{c}}^{\mathrm{HC}}\left(0\right) = 31.2(1)$~mT.

\subsection{Surface and upper critical field}
\label{UpperHc2}

\begin{figure}[t]
\centering
\includegraphics[width=0.7\columnwidth]{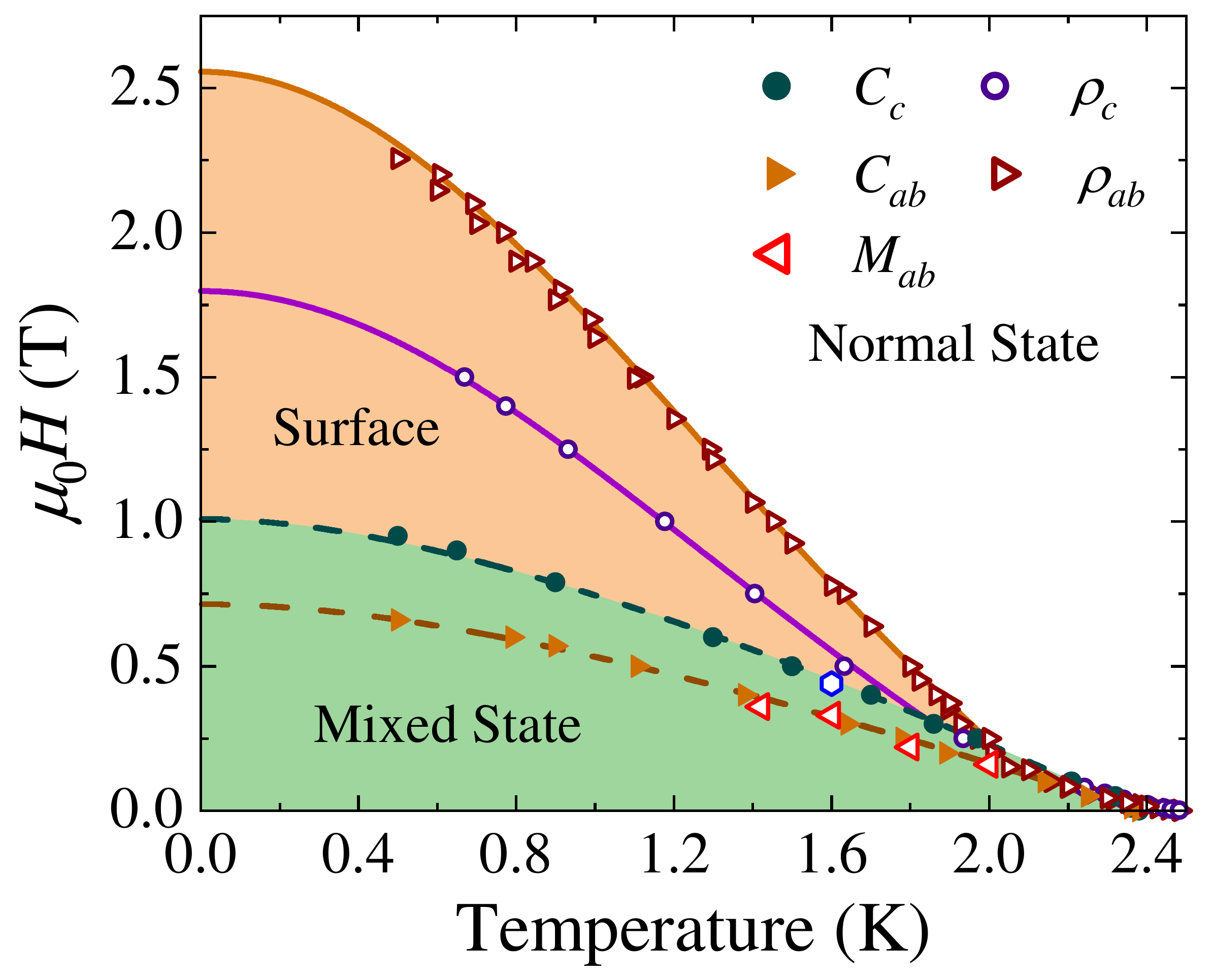}
\caption{(Color online) (a) Phase diagram showing the temperature dependence of the critical fields determined from heat capacity, $C$, the resistivity $\rho$ with a magnetic field applied in the $ab$ plane or along the $c$ axis, and five points from magnetization versus field curves. Dashed lines show the upper critical fields $H_{\mathrm{c2}}\left(T\right)$, obtained from fits made using the WHH model with $\mu_{0}H_{\mathrm{c2}}^{ab}\left(0\right) = 0.71(3)$~T and $\mu_{0}H_{\mathrm{c2}}^{c}\left(0\right) = 1.01(5)$~T. Solids lines mark $H_{\mathrm{c3}}\left(T\right)$ estimated using a GL model. }
\label{FIG: La7Ir3 UCF}
\end{figure}

The upper critical field, $H_{\mathrm{c2}}$ and its anisotropy, can give clues to the superconducting pairing mechanism and the nature of the superconducting gap. Critical field values were estimated by taking the midpoint of the superconducting transition in both the resistivity [see Fig.~\ref{FIG: La7Ir3 Resistivity}(b)] and heat capacity [see Fig.~\ref{FIG: La7Ir3 HC}(a)] versus temperature data, as well as from a distinct change in the slope in the $M(H)$ curves. The $H$-$T$ phase diagram obtained is shown in Fig.~\ref{FIG: La7Ir3 UCF}. There is a noticeable anisotropy in the upper critical field determined from heat capacity measurements. Similar measurements performed on single crystals of La$_{7}$Ni$_{3}$ show no difference between the different crystallographic directions~\cite{Nakamura17}.

The upper critical fields obtained from the heat capacity with a magnetic field applied along the two orthogonal crystallographic directions can be fit using the Werthamer-Helfand-Hohenberg (WHH) model~\cite{WHH}. The WHH model has proved a useful tool in predicting the upper critical field values for intermetallic superconductors, including those with a NCS structure, as it allows for the inclusion of the effects of spin-orbit coupling and Pauli spin paramagnetism. When the magnetic field is applied along the $[2\bar{1}\bar{1}0]$ direction (in the $ab$ plane) the gradient near $T_{\mathrm{c}}$, $\mu_{0}dH_{\mathrm{c2}}^{ab}\left(T\right)/dT\vert_{T=T_{\mathrm{c}}} = -0.445(5)$~T/K. We can then determine the orbital critical field to be $\mu_{0}H^{\mathrm{orb}}_{\mathrm{c2}}\left(0\right) = 0.73(1)$~T and the upper critical field to be $\mu_{0}H_{\mathrm{c2}}^{ab}\left(0\right) = 0.71(3)$~T. Similarly, for fields applied along the $c$ axis of La$_{7}$Ir$_{3}$,  $\mu_{0}dH_{\mathrm{c2}}^c\left(T\right)/dT\vert_{T=T_{\mathrm{c}}} = -0.62(1)$~T/K giving a $\mu_{0}H^{\mathrm{orb}}_{\mathrm{c2}}\left(0\right) = 1.08(2)$~T and $\mu_{0}H_{\mathrm{c2}}^c\left(0\right) = 1.01(5)$~T. The Pauli limiting field is much larger than these upper critical fields with $\mu_{0}H^{\mathrm{Pauli}}_{\mathrm{c2}}\left(0\right) = 4.4(1)$~T.

Using the $H_{\mathrm{c2}}(0)$ and $H_{\mathrm{c1}}(0)$ values parallel and perpendicular to the $c$ axis determined from the heat capacity and magnetization data, the Ginzburg-Landau coherence length, $\xi$, penetration depth, $\lambda$, and Ginzburg-Landau parameter, $\kappa_{\mathrm{GL}} = \lambda/\xi$, can be calculated as shown in the Appendix. The thermodynamic critical field can then be estimated using $H_{\mathrm{c}} = \Phi_{0}/2\sqrt{2}\pi\xi\lambda$ where $\Phi_{0}$ is a magnetic flux quantum giving an $H_{\mathrm{c}}$ between 34.6 and 37.5~mT. This is in fair agreement with the 31.2~mT, determined from the zero-field heat capacity and indicates that the $H_{\mathrm{c2}}(0)$ values determined above are the bulk values for La$_{7}$Ir$_{3}$.

The critical fields determined from the resistivity data, $H_{\mathrm{c}}^{\mathrm{\rho}} $ agree with those obtained from heat capacity data at low field. However, the normal-superconducting phase boundary determined from the resistivity data shows a clear change in gradient below $\sim2$~K. Due to the unusual shape of this curve, fits made using standard models cannot reproduce the behavior. However, a rough estimate of $H_{\mathrm{c}}^{\rho}\left(0\right)$ using the Ginzburg-Landau model $H_{\mathrm{c2}}\left(T\right) = H_{\mathrm{c2}}\left(0\right)\left[\frac{1 - \left(T/T_{\mathrm{c}}\right)^{2}}{1 + \left(T/T_{\mathrm{c}}\right)^{2}}\right]$ and the data below 1.8~K give critical fields $H_{\mathrm{c}}^{\rho}\left(0\right)$ at 0~K of 2.5(1)~T for $H$ in the $ab$ plane and 1.8(1)~T for $H \parallel c$. These values are both above the $\mu_{0}H_{\mathrm{c2}}^{\mathrm{\rho}}\left(0\right)=1.6$~T reported from $\rho\left(T\right)$ measurements on polycrystalline samples of La$_{7}$Ir$_{3}$~\cite{Li17}, and is also significantly higher than the $H_{\mathrm{c2}}\left(0\right)$ values obtained here from heat capacity and magnetization measurements. Any enhancement in the upper critical field seen from resistivity is unlikely to be due to impurities given the evidence for the high-quality of the single crystals. This behavior is evidence for surface superconductivity, with the transition in resistivity tracing the temperature dependence of $H_{\mathrm{c3}}$, a suggestion that is supported by the noticeable broadening of the superconducting transition in $\rho\left(T\right)$ when the applied field is increased~\cite{Zeinali16}.

An enhancement in the upper critical fields obtained from resistivity measurements is also seen in several other NCS superconductors including BeAu~\cite{Rebar19}, BiPd~\cite{Peets16}, LaRhSi$_{3}$~\cite{Kimura16}, BaPtSi$_{3}$~\cite{Bauer09}, LuNiSi$_{3}$~\cite{Arantes19}, YNiSi$_{3}$~\cite{Arantes19}, LaPdSi$_{3}$~\cite{Smidman14} and LaPtSi$_{3}$~\cite{Smidman14}. The materials listed above differ from La$_{7}$Ir$_{3}$ as they are type-I or marginal type I/II superconductors, with upper critical fields between 10 and 100's of mT, whereas La$_{7}$Ir$_{3}$ is clearly type-II with a bulk critical field, $H_{\mathrm{c2}}(0)$, an order of magnitude higher.

Assuming the upper critical field determined from resistivity is the surface critical field, i.e. $H_{\mathrm{c2}}^{\mathrm{\rho}}(0) = H_{\mathrm{c3}}(0)$, then the ratio $H_{\mathrm{c3}}(0)/H_{\mathrm{c2}}(0)$, is 1.8(1) and 3.5(1) for $H$ applied parallel and perpendicular to $c$, respectively~\cite{RatioNote}. Both values, but especially the latter, are higher than the Saint-James and de Gennes limit for type-II superconductors where $H_{\mathrm{c3}} = 1.695 H_{\mathrm{c2}}$~\cite{Saint-James64}. The validity of this ratio is rather limited and can be affected by many factors such as an intrinsic anisotropy~\cite{Kogan02}, sample shape~\cite{Moshchalkov95}, and impurities or disorder that determine the clean and the dirty limits~\cite{Gorokhov05}.

It has been proposed that local critical field can be enhanced by magnetoelectric effects at twin boundaries in NCS superconductors, however, due to the single crystal nature of this sample we would expect this enhancement to be small~\cite{Aoyama14}. In MgB$_{2}$ a $H_{\mathrm{c3}}/H_{\mathrm{c2}} > 1.695$ has been associated with two-gap superconductivity. From the heat capacity data we see no evidence of two-gap superconductivity in La$_{7}$Ir$_{3}$, however, a precise knowledge of the electronic heat capacity  below 10\% of $T_{\mathrm{c}}$ ($T<240$~mK) is required to reliably determine the nature of the superconducting gap. It may be possible for La$_{7}$Ir$_{3}$ to have two nodeless gaps with similar magnitudes which would be difficult to identify from heat capacity data alone~\cite{Frigeri06}.

We also note that the divergence of the upper critical field determined from resistivity corresponds to the onset temperature ($T\approx 2$~K) of the TRSB reported in polycrystalline samples of La$_{7}$Ir$_{3}$~\cite{La7Ir3}. This could suggest that the mechanism behind the TRSB may be connected with the enhancement in the critical field La$_{7}$Ir$_{3}$ although no reports of a similar behavior in $H_{\mathrm{c2}}(T)$ - $H_{\mathrm{c3}}(T)$ have been reported in other members of La$_{7}X_{3}$ family ($X=$~Rh, Pd or Ni)~\cite{Pedrazzini00,La7Rh3,Mayoh21,Nakamura17} Note, however, there have been similar reports for an enhanced critical field as determined from resistivity measurements in LaNiC$_2$, with a similar enhancement $H_{\mathrm{c3}}(0)/H_{\mathrm{c2}}(0) \approx 3$, although surface superconductivity was not explicitly discussed~\cite{Hirose2012}.

%%%%%%%%%%%%%%%%%%%%%%%%%%%%%%%%%%%%%%%%%%

\section{Summary}

We have successfully grown a large high-quality single crystal of La$_{7}$Ir$_{3}$ by the Czochralski process using a radio frequency furnace. Powder x-ray diffraction and EDX have confirmed the structure and stoichiometry of the single crystals, while Laue patterns show sharp reciprocal space lattices indicative of a high quality single crystal. La$_{7}$Ir$_{3}$ exhibits a superconducting transition at $T_{\mathrm{c}} = 2.38(5)$~K which is higher than reported in polycrystalline samples. Initial measurements indicate an anisotropy in both the lower and upper critical fields. The critical field determined from the resistivity data is found to be much more robust, persisting to higher magnetic fields than its counterpart determined from heat capacity measurements, providing strong evidence for a previously unreported existence of a surface critical field in La$_{7}$Ir$_{3}$. Further investigations on these crystals of the temperature dependence of the superconducting penetration depth using the tunnel-diode oscillator technique or muon spectroscopy will be important in determining the superconducting order parameter, as well as whether TRSB is observed in single crystal La$_{7}$Ir$_{3}$.

\appendix*

\section{Anisotropic superconducting parameters}\label{APD: superconducting parameters}

The anisotropy in the upper and lower critical fields between the two crystallographic directions indicates there must be an anisotropy in the effective mass of the carriers and the flux-line cores. This must be considered when calculating the relevant superconducting length scales~\cite{Poole}. Assuming La$_{7}$Ir$_{3}$ is axially symmetric about the $c$ axis, the Ginzburg-Landau coherence length can be calculated using $H_{\mathrm{c2}}^{ab} = \Phi_{0}/2\pi\xi_{\mathrm{ab}}\xi_{\mathrm{c}}$ and $H_{\mathrm{c2}}^{c} = \Phi_{0}/2\pi\xi_{\mathrm{ab}}^{2}$ to give $\xi_{\mathrm{c}}(0) = 25.6(6)$~nm and $\xi_{\mathrm{ab}}(0) = 18.1(1)$~nm. To calculate the penetration depth, $\lambda$, and the GL parameter, $\kappa$, the equations  $H_{\mathrm{c1}}^{ab}=\Phi_{0}\ln\kappa_{\mathrm{ab}}/4\pi\lambda_{\mathrm{ab}}\lambda_{\mathrm{c}}$, $H_{\mathrm{c1}}^{c} = \Phi_{0}\ln\kappa_{\mathrm{c}}/4\pi\lambda_{\mathrm{ab}}^{2}$,  $\kappa_{\mathrm{ab}} = |\lambda_{\mathrm{ab}}\lambda_{\mathrm{c}}/\xi_{\mathrm{ab}}\xi_{\mathrm{c}}|^{1/2}$ and $\kappa_{\mathrm{c}} = \lambda_{\mathrm{ab}}/\xi_{\mathrm{ab}}$ are solved simultaneously to give $\lambda_{\mathrm{ab}}(0) = 372(2)$~nm, $\lambda_{\mathrm{c}}(0) = 241(2)$~nm, $\kappa_{\mathrm{ab}} = 13.9$ and $\kappa_{\mathrm{c}} = 20.6$.

\begin{acknowledgments}
We would like to acknowledge Ali Julian, Tom Orton and Patrick Ruddy for their technical support. We would also like to thank Steve York for their assistance with the energy-dispersive x-ray spectroscopy measurements. This work was financially supported by two Engineering and Physical Sciences Research Council grants: EP/M028771/1 and EP/T005963/1. 
\end{acknowledgments}

\bibliography{La7Ir3_DM_References}

\begin{thebibliography}{50}
\bibitem [{Rat()}]{RatioNote}%
  \BibitemOpen
  \href@noop {} {}\bibinfo {note} {Similar values for the ratio
  $H_{\mathrm{c3}}/H_{\mathrm{c2}}$ are obtained using the
  $H_{\mathrm{c3}}^c(T)$ and $H_{\mathrm{c2}}^2(T)$ values at the lowest
  temperature available ($\sim0.6$~K).}\BibitemShut {Stop}%
\end{thebibliography}

%--------------------- Figures ------------

\begin{table}[t]
\caption{Crystallographic parameters for La$_{7}$Ir$_{3}$ obtained from a Rietveld refinement of the powder x-ray diffraction data collected at room temperature.} \label{Tab:La7Ir3 pxrd}
\begin{center}
\begin{ruledtabular}

\begin{tabular}{llllll} 
  \multicolumn{3}{l}{Structure} & \multicolumn{3}{l}{Th$_7$Fe$_3$} \\ 
  \multicolumn{3}{l}{} & \multicolumn{3}{l}{Hexagonal} \\
  \multicolumn{3}{l}{Space group} & \multicolumn{3}{l}{$P 6_{3} mc$ (No. 186)} \\
  \multicolumn{3}{l}{Formula units/unit cell ($Z$)} & \multicolumn{3}{l}{2} \\
  \multicolumn{3}{l}{Lattice parameter} & \multicolumn{3}{l}{}\\
  \multicolumn{3}{l}{$a$ (\AA)} & \multicolumn{3}{l}{10.2382(5)} \\
  \multicolumn{3}{l}{$c$ (\AA)}  & \multicolumn{3}{l}{6.4778(5)} \\
  \multicolumn{3}{l}{$V_{\mathrm{cell}}$ (\AA$^3$)} & \multicolumn{3}{l}{588.04(8)} \\
  \multicolumn{3}{l}{$\rho$ (g/cm$^{3}$)} & \multicolumn{3}{l}{8.747(1)} \\
  \multicolumn{3}{l}{$R_{\mathrm{wp}}$ (\%)} & \multicolumn{3}{l}{6.92} \\
  \multicolumn{3}{l}{$R_{\mathrm{exp}}$ (\%)} & \multicolumn{3}{l}{0.98}\\
  \multicolumn{3}{l}{$R_{\mathrm{Bragg}}$ (\%)} & \multicolumn{3}{l}{3.39} \\
  \multicolumn{3}{l}{$\chi^{2}$} & \multicolumn{3}{l}{50.9} \\
 \hline
Atom & Wyckoff & Occupancy & x & y & z  \\
& Position &&& \\
 \hline
Ir1 & 6c & 1 & 0.8125(1) & 0.1875(1) & 0.3185(5)  \\
La1 & 2b & 1 & 1/3 & 2/3 & 0.0545(10) \\
La2 & 6c & 1 & 0.1256(2) & 0.8744(2) & 0.2717(3) \\
La3 & 6c & 1 & 0.5388(2) & 0.4612(2) & 0.0726(4) \\
\end{tabular}
%\end{tabular}
\end{ruledtabular}
\end{center}
\end{table}

\newpage

\end{document}

\renewcommand{\thefigure}{S\arabic{figure}}
\renewcommand{\thetable}{S\arabic{table}}

\title{Supplemental Material: Anisotropic superconductivity and unusually robust electronic critical field in single crystal La$_{7}$Ir$_{3}$}

\author{D. A. Mayoh}
\email[]{d.mayoh.1@warwick.ac.uk}
\affiliation{Physics Department, University of Warwick, Coventry, CV4 7AL, United Kingdom}

\author{S. J. R. Holt}
\affiliation{Physics Department, University of Warwick, Coventry, CV4 7AL, United Kingdom}

\author{T. Takabatake}
\affiliation{Graduate School of Advanced Sciences of Matter, Hiroshima University, Higashi-Hiroshima 739-8530, Japan}

\author{G. Balakrishnan}
\affiliation{Physics Department, University of Warwick, Coventry, CV4 7AL, United Kingdom}

\author{M. R. Lees}
\email[]{m.r.lees@warwick.ac.uk}
\affiliation{Physics Department, University of Warwick, Coventry, CV4 7AL, United Kingdom}

\maketitle

\section{Energy dispersive x-ray analysis}

\begin{figure}[h!]
\centering
\includegraphics[width=\columnwidth]{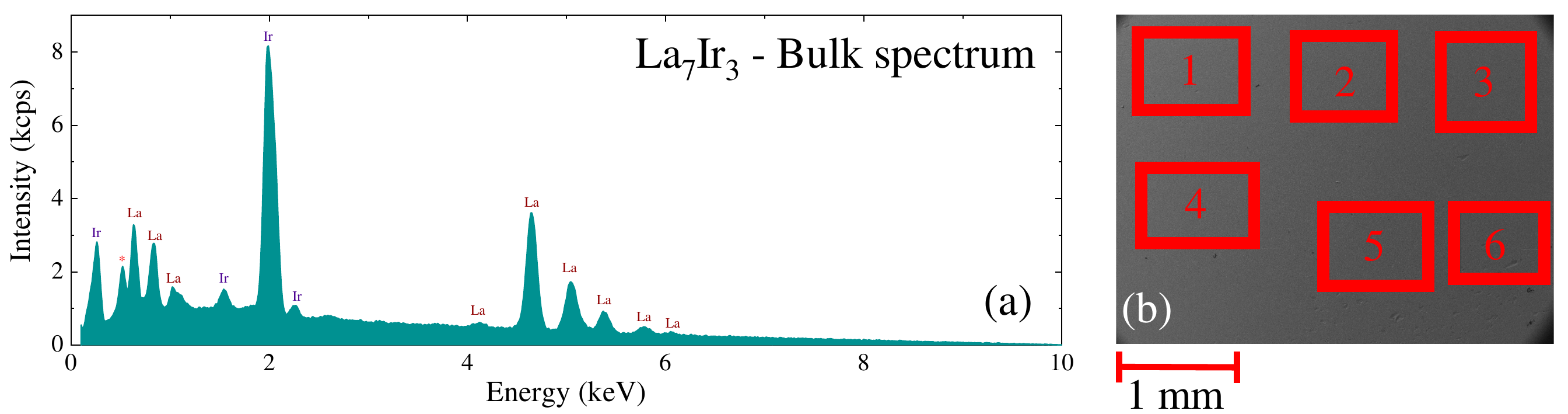}
\caption{(a) EDX spectrum for a single crystal of La$_{7}$Ir$_{3}$ where the intensities and energies of the peaks allow the elemental composition of the sample to be determined. The bulk stoichiometry was measured to be La$_{7.05(2)}$Ir$_{2.95(2)}$. The red star indicates an oxygen peak. (b) Scanning electron microscope image of the surface of the single crystal studied. EDX spectra were collected over the whole of the crystal as well as within the marked boxes.}
\label{FIG: La7Ir3 EDX}
\end{figure}

A ZEISS GeminiSEM 500 was used to perform energy-dispersive x-ray spectroscopy (EDX) measurements on an isolated piece of the La$_{7}$Ir$_{3}$ boule, to check the stoichiometry and search for any elemental variations across the sample [see Fig.~\ref{FIG: La7Ir3 EDX}(a)]. Measurements were made averaging across the bulk of the sample as well as at six selected sites on the crystal surface [see Fig.~\ref{FIG: La7Ir3 EDX}(b)]. The average stoichiometry was measured to be 70.5(2)\% La and 29.5(2)\% Ir. Table~\ref{Tab:La7Ir3 EDX} shows the relative elemental quanties at each site measured and across the bulk of the sample.

\section{Magnetic Susceptibility}

A Quantum Design Magnetic Property Measurement System was used to perform ac and dc susceptibility measurements. The dc susceptibility versus temperature, $\chi_{\mathrm{dc}}\left(T\right)$, measurements were made in zero-field-cooled-warming (ZFCW) and field-cooled-cooling (FCC) modes between 1.8 and 2.8~K in applied fields of up to 0.3~T. The ac susceptibility versus temperature, $\chi_{\mathrm{ac}}\left(T\right)$, data were collected on warming between 1.8 and 2.8~K using an ac field of 0.3~mT with a frequency of 3~Hz in dc applied fields of up to 10~mT. All the magnetic susceptibility data were corrected for the demagnetization factors determined from the sample shapes~\cite{Aharoni1998}. 

\subsection{DC Susceptibility}

\begin{figure}[h]
\centering
\includegraphics[width=0.6\columnwidth]{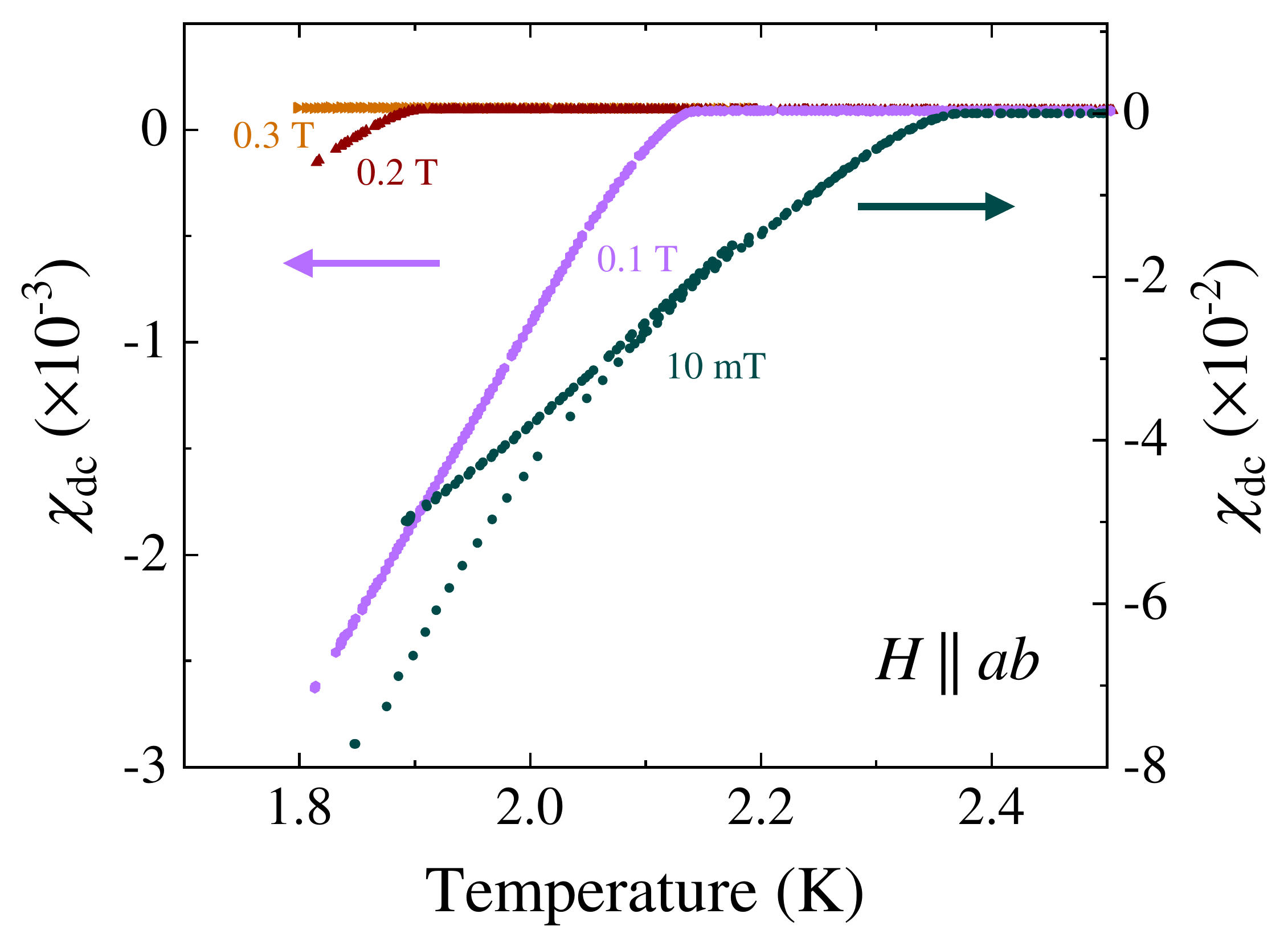}
\caption{Temperature dependence of the dc susceptibility, $\chi_{\mathrm{dc}}\left(T\right)$, of La$_{7}$Ir$_{3}$ in different dc fields applied in the $ab$ plane. Note, the feature in the data at $\sim 2.2$~K is due to a temperature instability in the magnetometer sweeping through the $\lambda$-point of helium.}
\label{FIG: La7Ir3 DC sus SM}
\end{figure}

The temperature dependence of $\chi_{\mathrm{dc}}\left(T\right)$ in 1~mT is shown in Fig.~3(a) of the main text. The field dependence of the $\chi_{\mathrm{dc}}\left(T\right)$ in higher fields is shown in Fig.~\ref{FIG: La7Ir3 DC sus SM}. In 10~mT, $T_{\mathrm{c}} = 2.36(5)$~K, in good agreement with both the heat capacity and resistivity versus temperature data. Immediately below $T_{\mathrm{c}}$, $\chi_{\mathrm{dc}}\left(T\right)$ is reversible until $\sim 2.1$~K with a small amount of hysteresis between the ZFCW and FCC curves at lower temperature. A dc field of 10~mT is above $H_{\mathrm{c1}}\left(0\right)$ and so the material is in the mixed state, with $\chi_{\mathrm{dc}}$ less than 10\% of the full Meissner fraction at 1.8~K. 

In higher fields, $\chi_{\mathrm{dc}}\left(T\right)$ is reversible. In 0.1~T and above $\left|\chi_{\mathrm{dc}}\right|<<1$, indicating considerable flux penetration in this weakly pinning system. At 0.3~T, the $T_{\mathrm{c}}$ is below 1.8~K, in agreement with the heat capacity data, but in clear contradiction with resistivity versus temperature and resistivity versus field data that indicate superconductivity persists to higher fields. For example, the transition at 1.8~K in resistivity versus field data is above 0.3~T (see Fig.~\ref{FIG: La7Ir3 Resistivity SM}). This highlights that the dc susceptibility data probe the bulk superconductivity at these fields.

\subsection{AC Susceptibility}

$\chi_{\mathrm{ac}}$ versus temperature data were collected in several dc fields applied in the $ab$ plane. In zero dc field (the trapped field of the superconducting solenoid after quenching) the superconducting transition at $2.40(5)$~K is sharp with no features. This underlines the single-phase nature of the sample. In a dc field of 0.4~mT, a full Meissner fraction is still observed at lower temperature, which is consistent with the  $H_{\mathrm{c1}}\left(T\right)$ curve shown in Fig.~3(c) of the main text. At higher dc fields, in the mixed state, the in-phase component of the ac susceptibility $\left|\chi'_{\mathrm{ac}}\right|$ is smaller than 1 over the entire temperature window, as expected for a sample in the mixed state. The large out-of-phase component $\chi''_{\mathrm{ac}}$ shows that there are considerable losses in the sample, consistent with flux motion (weak pinning) in the mixed state.

\newpage
\section{Electrical Resistivity}

Electrical resistivity versus temperature curves for a single crystal of La$_{7}$Ir$_{3}$  in selected applied fields between 0 and 1.5~T are shown in Fig.~\ref{FIG: La7Ir3 Resistivity SM}(a). The field was applied along the $[0001]$ direction and the current passed along the $[\bar{1}2\bar{1}0]$ direction, within the $ab$ plane. Electrical resistivity versus applied magnetic field measurements for a single crystal of La$_{7}$Ir$_{3}$ at selected temperatures between 0.5 and 2.4~K are shown in Fig.~\ref{FIG: La7Ir3 Resistivity SM}(b). The field was applied along the$[2\bar{1}\bar{1}0]$ direction and the current passed along the $[\bar{1}2\bar{1}0]$ direction, within the $ab$ plane.

\bibliography{La7Ir3_DM_References}

%------ Figures -------

\begin{table}[h]
\setlength{\tabcolsep}{18pt}
\caption{Composition of a single crystal of La$_{7}$Ir$_{3}$ determined from energy dispersive x-ray analysis.} \label{Tab:La7Ir3 EDX}
\begin{center}

\begin{tabular}{llll} 

\hline
\hline
Spectrum & La & Ir\\
Label & (\%) & (\%) \\
 \hline
Bulk & 70.5(2) & 29.5(2) \\
1 & 70.6(2) & 29.4(2)\\
2 & 70.6(2) & 29.4(2) \\
3 & 70.9(2) & 29.1(2) \\
4 & 70.8(2) & 29.5(2) \\
5 & 70.8(2) & 29.5(2) \\
6 & 71.2(2) & 28.8(2) \\
\hline
\end{tabular}
\end{center}
\end{table}

\newpage

\begin{figure}[h]
\centering
\includegraphics[width=0.6\columnwidth]{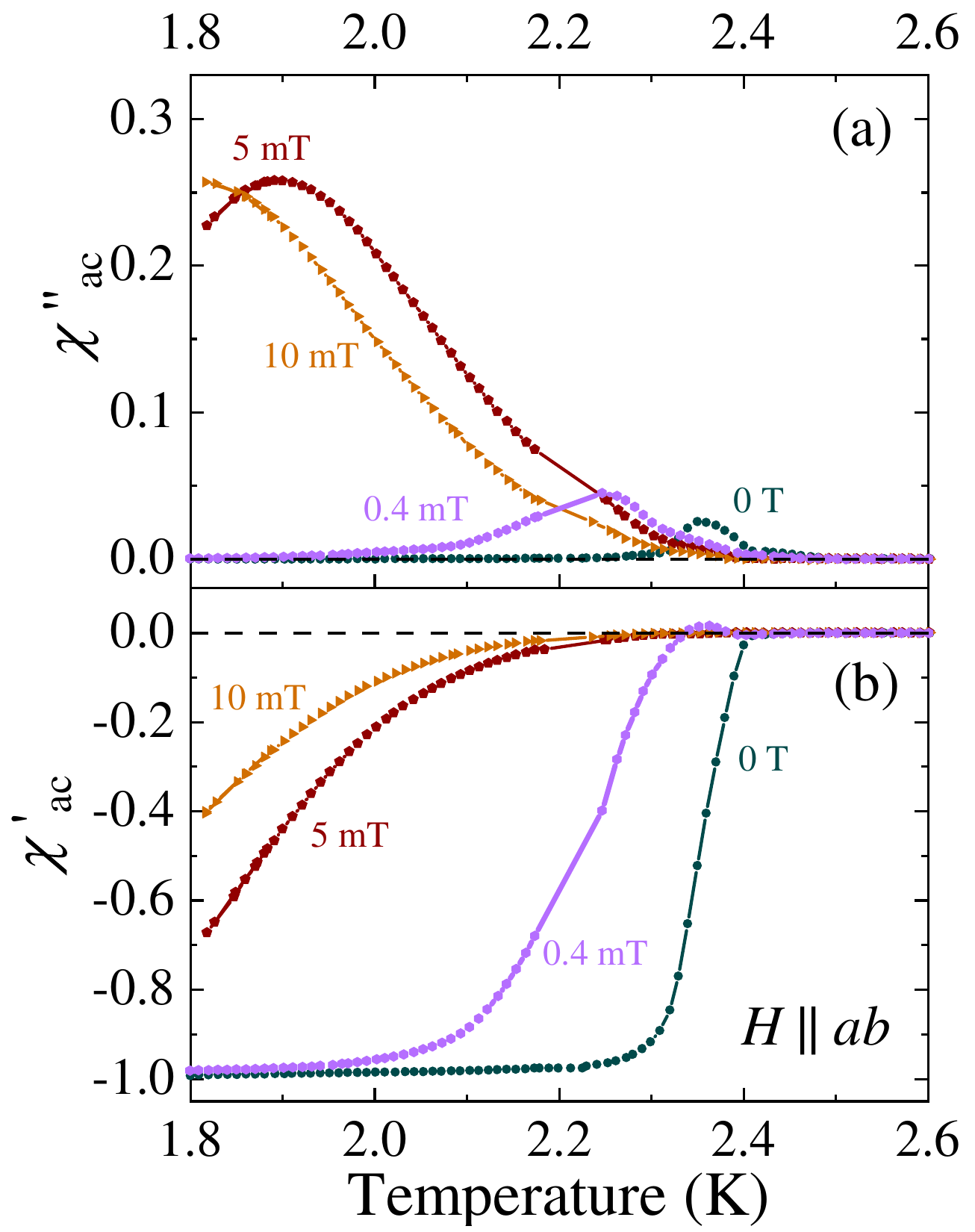}
\caption{ (a) Imaginary part $\chi''_{\mathrm{ac}}$ and (b) real part $\chi'_{\mathrm{ac}}$ of the ac susceptibility versus temperature of La$_{7}$Ir$_{3}$ collected in an ac field of 0.3~mT and different dc fields applied in the $ab$ plane. A sharp superconducting transition is observed in zero dc field at $2.40(5)$~K. Note, the features in the data at $\sim 2.2$~K are due to a temperature instability in the magnetometer around the $\lambda$-point of helium.}
\label{FIG: La7Ir3 AC sus SM}
\end{figure}

\begin{figure}[h]
\centering
\includegraphics[width=\columnwidth]{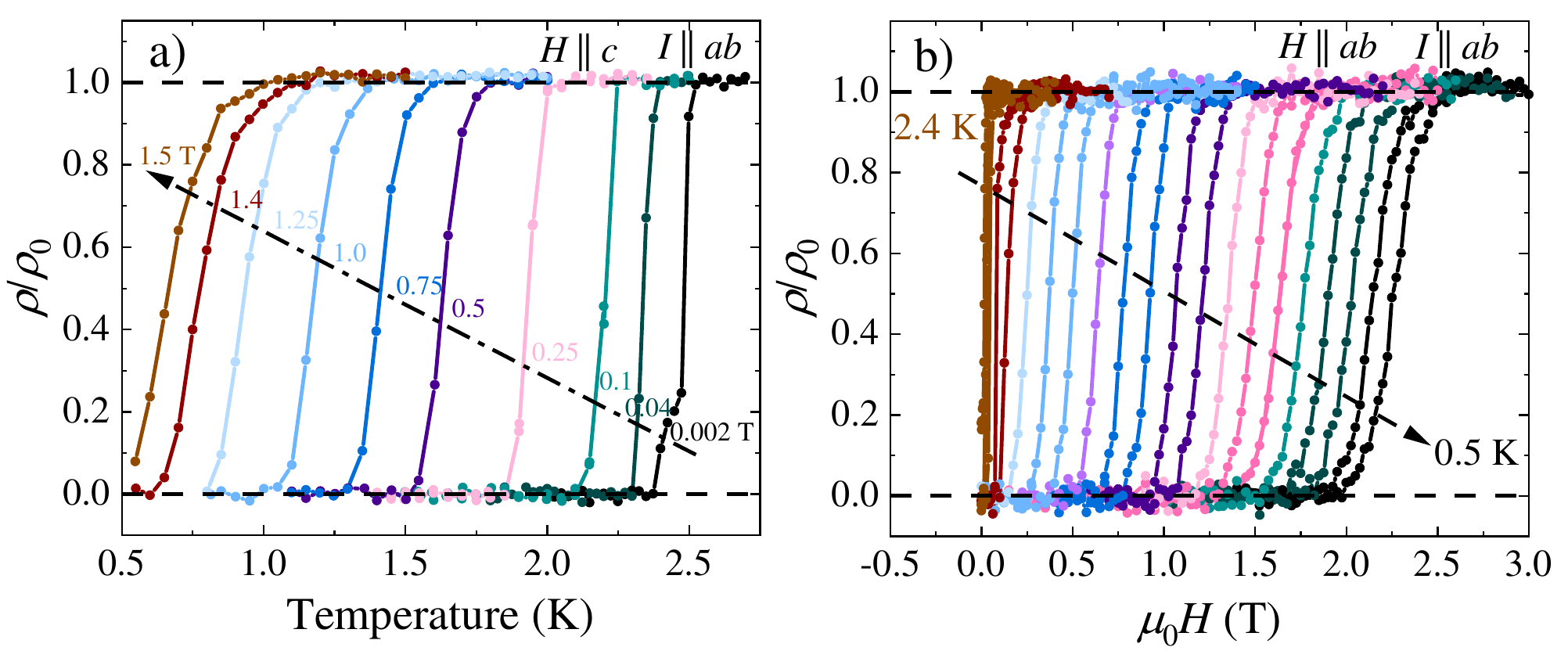}
\caption{(a) Electrical resistivity versus temperature for a single crystal of La$_{7}$Ir$_{3}$ in selected applied fields between 0 and 1.5~T. The field was applied along the $[0001]$ direction and the current passed along the $[\bar{1}2\bar{1}0]$ direction, within the $ab$ plane. (b) Electrical resistivity versus applied magnetic field for a single crystal of La$_{7}$Ir$_{3}$ at selected temperatures between 0.5 and 2.4~K. The field was applied along the $[2\bar{1}\bar{1}0]$ direction and the current passed along the $[\bar{1}2\bar{1}0]$ direction, within the $ab$ plane.}
\label{FIG: La7Ir3 Resistivity SM}
\end{figure}